\documentclass[conference]{IEEEtran}
\IEEEoverridecommandlockouts

\usepackage{cite}
\usepackage{amsmath,amssymb,amsfonts}
\usepackage{graphicx}
\usepackage{textcomp}
\usepackage{xcolor}
\def\BibTeX{{\rm B\kern-.05em{\sc i\kern-.025em b}\kern-.08em
    T\kern-.1667em\lower.7ex\hbox{E}\kern-.125emX}}

\usepackage{booktabs}
\usepackage{multirow}
\usepackage{algorithm}
\usepackage{algpseudocode}
\usepackage{graphicx}
\usepackage{gensymb}
\usepackage[table,xcdraw]{xcolor}
\usepackage{colortbl}
\usepackage{relsize}
\usepackage[normalem]{ulem}
\usepackage{subcaption}
\usepackage{enumitem}
\setlist{itemsep=1pt, parsep=0pt}
\useunder{\uline}{\ul}{}

\algnewcommand\algorithmicforeach{\textbf{for each}}
\algdef{SE}[FOREACH]{ForEach}{EndForEach}[1]
  {\algorithmicforeach\ #1\ \textbf{do}}
  {\algorithmicend\ \algorithmicforeach}
\algtext*{EndForEach}  

\begin{document}

\title{A Subjective Study on a New Sharpness Informed Class of Metrics \\
\thanks{This work was supported in part by a Google / YouTube Faculty award.}
}

\author{\IEEEauthorblockN{Uditangshu Aurangabadkar, Vibhoothi Vibhoothi, Darren Ramsook, Anil Kokaram}
\IEEEauthorblockA{{Sigmedia Group, Department of Electronic and Electrical Engineering} \\
\textit{Trinity College Dublin}, Dublin, Ireland \\
\{aurangau, vibhootv, dramsook, anil.kokaram\}@tcd.ie}}

\maketitle

\begin{abstract}
Perceptual loss functions in Deep Neural Network (DNN) deblurring architectures improve the overall quality of restored images. However, few focus on explicitly targeting sharpness in the restorations. We conduct a subjective study of models trained with and without losses which explicitly target sharpness using a four-protocol approach, exploring preferred sharpness levels and effects on image quality. We introduce a novel dataset of images with uniform sharpness increments along with Difference Mean Opinion Scores (DMOS). Additionally, we propose a novel class of \textit{Sharpness Informed} (SI) Image Quality Assessment (IQA) metrics which properly penalize over-sharpening. Our new \text{SI-PSNR} metric outperforms all other PSNR variants in terms of correlation statistics on IQA benchmarking datasets. We show that, on average, images restored using a sharpness-aware composite loss are preferred in $67\%$ of binarized comparisons, as opposed to losses that do not explicitly target sharpness.
\end{abstract}


\section{Introduction}
\label{sec:intro}
Over the past few years, deblurring has seen a paradigm shift from traditional methods to Deep Neural Network (DNN) based architectures~\cite{kupyn2018deblurgan} driven by their fast inference times and the capability to handle complex blur kernels. Beyond designing complex architectures, works~\cite{he2024dual} have begun to focus on perceptual losses to produce restorations that align well with the Human Visual System (HVS).  

Sharpness is important for perceived image quality, yet few DNN-based image deblurring models explicitly target sharpness as part of the loss function. A key challenge in optimizing DNNs for sharpness is that there are selectively few perceptual measures that properly take into account sharpness and correctly penalize ringing. Although metrics such as SSIM~\cite{1284395}, PSNR-HVS~\cite{egiazarian2006new} and LPIPS~\cite{zhang2018unreasonable} measure overall image quality, they fail to distinguish positive sharpness improvements from ringing. An example is the loss introduced by Aurangabadkar et al.~\cite{aurangabadkar2024sharpness}, which enhances sharpness based on a no-reference metric $Q$ introduced by Zhu and Milanfar~\cite{zhu2010automatic}. While some works such as~\cite{zamir2021multi} use edge-losses for sharpening restorations, very few have assessed the perceptual quality of restored images using hybrid or sharpness losses. 

In the current work, we first demonstrate that subjects prefer sharper images as compared to original reference images, suggesting that deblurring models should produce slightly sharper restorations as compared to Ground Truth (GT). We then perform a subjective study and assess whether using sharpness loss $Q$ produces visually sharper and aesthetically better images. We show that most Image Quality Assessment (IQA) metrics do not correlate well with perceptual scores for \textit{sharpened} and \textit{over-sharpened} images. Therefore, we propose a class of \textit{Sharpness Informed} (SI) metrics which correctly penalize over-sharpness and demonstrate how they can be integrated into classical measures such as Peak Signal-to-Noise Ratio (PSNR) or SSIM. The contributions of this work are as follows.

\begin{itemize}
    \item A subjective study comparing DNN architectures with and without losses that explicitly target sharpness.
    \item A novel class of \textit{Sharpness Informed} metrics which correctly penalize ringing artifacts in an image.
    \item A novel dataset of images that have been sharpened with varying amounts of sharpness, along with their corresponding DMOS for benchmarking IQA metrics.
    \item Concrete evidence to show that people prefer slightly over-sharpened images as part of the viewing experience.
\end{itemize}

\section{Background}
\label{sec:backrgound}

Sharpness in images is a key characteristic that distinguishes visually better images. As described in the work by Krasula et al.~\cite{krasula2017quality}, every person has an acceptable threshold of sharpness. Moreover, sharpness is a content-specific property of the image, i.e. images with textural information (edges, corners) will have a different acceptable threshold than those with little to no textural information (flat areas). Although metrics such as BRSIQUE~\cite{mittal2012no} or $\Omega$~\cite{11226201} have thresholds of acceptable sharpness (extremely small increments in sharpness result in better scores), they suffer from a lack of interpretability.  Hence, it is a challenging task to design IQA metrics which correctly reward a positive sharpness gain, while penalizing ringing and halo artifacts. Most IQA benchmarking datasets such as TID2013~\cite{ponomarenko2015image} or LIVE~\cite{sheikh2006statistical} lack sharpened and over-sharpened images, which in turn make it difficult to calibrate measures aimed to assess such degradations. Therefore, we require an interpretable and bounded class of \textit{Sharpness Informed} metrics which properly penalize ringing and correlate well with HVS. 

Deblurring algorithms focus on restoring degraded images to their GT counterpart, such that the textural information is correctly restored using Mean Squared Error (MSE), Mean Absolute Error (MAE) and even composite losses containing perceptual metrics such as LPIPS. However, they do not explicitly sharpen the images, leading to a secondary sharpening process.

It is worth noting that few works perform subjective studies on images restored by State-of-the-Art (SOTA) deblurring models, more specifically whether using certain losses creates sharper and aesthetically pleasing restorations. 

\section{Sharpness Informed Metrics}
\label{sec:si_metrics}

$Q$ is a good indicator of sharpness in an image, but it suffers from limited interpretability when used as a no-reference (NR) metric. A high value of $Q$ either corresponds to a sharp image or an image with ringing. Classical metrics such as PSNR or SSIM remain popular for predicting image quality owing to their low complexity, but do not take into account sharpness or ringing. Therefore, we wish to incorporate sharpness information into these standard metrics. 

The intuition behind such a metric is to penalize areas with heavy ringing more than those with little to no ringing. We start by measuring the patch-wise Ringing Detection Ratio ($\alpha$) introduced in work~\cite{11226201} for each patch from GT ($I$) and restored image ($\tilde{I}$) as follows.
\begin{equation}
    \alpha = \frac{|\tilde{Q} - Q|}{Q}
\end{equation}

\begin{figure}
    \centering
    \includegraphics[width=1.0\linewidth]{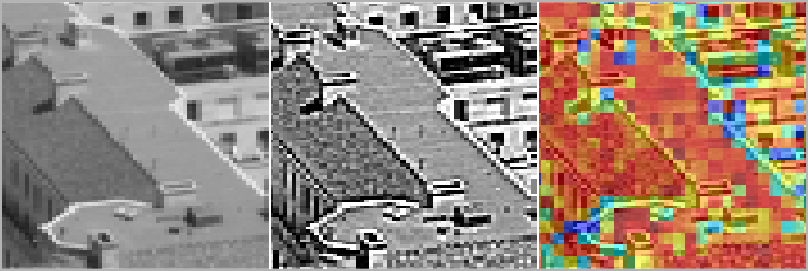}
    \caption{\textit{Ringing Map generated using Ringing Detection Ratio ($\alpha$) with a patch size of 4 $\times$ 4}. Left to Right: GT image, Over-sharpened image, Ringing Detection Heatmap. Areas in red (mostly edges) correspond to heavy ringing which are penalized more than areas in blue (flat patches).}
    \label{fig:ringing_detection_map}
\end{figure}

Essentially, $\alpha$ captures the difference in eigenvalue ratios of the gradients between $I$ and $\tilde{I}$. A higher value for a patch signifies stronger structural degradation, which in our case is ringing, though this holds true for blur and noise as well. Visual examples of this can be found in the supplementary materials\footnote{https://github.com/aurangau/MMSP2026}. We may use this ratio to generate a Ringing Map (Fig.~\ref{fig:ringing_detection_map}) where red indicates heavy ringing and blue indicates little to no ringing.

By weighting Mean Squared Error (MSE) with $\alpha$, we penalize areas with ringing. Exponentiating $\alpha$ to $\rho$ leads to better correlation with DMOS as it strengthens the penalizing factor. Details about our systematic ablation study can be found in our supplementary materials. Our Sharpness Informed (SI)-MSE is as follows.

\begin{equation}
    \text{SI-MSE}(I_K, \tilde{I}_K) = \text{MSE}(I_K, \tilde{I}_K) \cdot \alpha^\rho
    \label{eq:simse_formula}
\end{equation}

By averaging this metric over all patches, we then create a SI-PSNR for the whole picture as follows.

\begin{equation}
    \text{SI-PSNR}(I, \tilde{I}) = 10 \cdot \log_{10} \left ( \frac{255 ^2}{\text{SI-MSE}(I, \tilde{I})} \right )
    \label{eq:sipsnr_formula}
\end{equation}

\begin{algorithm}
\caption{Sharpness Informed (SI)--PSNR}
\begin{algorithmic}[1]
    \State \textbf{Input:} Images $I$, $\tilde{I}$
    \State \textbf{Output:} SI-PSNR($I$, $\tilde{I}$)
    \State Divide images $I$, $\tilde{I}$ into patches ($k \times k$)
    \ForEach{patch $j = 1$ to $N$}
        \State $\tilde{Q}_j = \mathbf{Q}(\tilde{I}_j)$
        \State $Q_j = \mathbf{Q}(I_j)$
        \State $\alpha_j = \dfrac{|\tilde{Q}_j - Q_j|}{Q_j}$
        \State $\text{SI-MSE}_j(\tilde{I}_j, I_j) = \text{MSE}_j(\tilde{I}_j, I_j) \cdot \alpha_{j}^{\rho}$
    \EndForEach                         
    \State $\text{SI-MSE}(\tilde{I}, I) = \dfrac{1}{N} \mathlarger{\sum_{j=1}^{N}} \text{SI-MSE}_j(\tilde{I}_j, I_j)$
    \State $\mathrm{SI\text{-}PSNR}(\tilde{I}, I)
        = 10 \cdot \log_{10}\!\left(\dfrac{255^2}{\mathrm{SI\text{-}MSE}(\tilde{I}, I)}\right)$
\end{algorithmic}
\label{alg:Q_blending}
\end{algorithm}

A similar treatment applied to SSIM yields SI-SSIM as follows.

\begin{equation}
    \text{SI-SSIM}(I, \tilde{I}) = \text{SSIM}(I, \tilde{I}) \cdot \alpha ^ \rho
    \label{eq:sissim_formula}
\end{equation}

\section{Experimental Setup}
\label{sec:setup}
All subjective experiments were conducted in a laboratory-compliant room with the recommendations of ITU-BT.500-15~\cite{bt2019methodologies}. Test images were displayed on a 32-inch Sony BVM-X300v2 4K OLED-critical reference monitor using a custom interactive interface developed in MATLAB. The viewing distance was $1.6$H and the background luminance was approximately $4.2$ nits. 

\subsection{Datasets, Deblurring Methods and Loss Functions}
\label{sec:datasets_methods}
We chose two Deep Neural Network (DNN)-based deblurring methods, namely ARKNet~\cite{aurangabadkar2024sharpness} and XY-Deblur~\cite{ji2022xydeblur}. ARKNet is a standard U-Net-based network which consists of $\approx$ 4.2 million trainable parameters and acts as our baseline model. XY-Deblur~\cite{ji2022xydeblur} ($\approx$ 4.9 Million trainable parameters) is a deblurring mechanism that was first introduced for removing motion blur from images. The model uses a dual-decoder mechanism and has been shown to produce superior results as compared to using only a single decoder. 

Images from DIV2K~\cite{Agustsson_2017_CVPR_Workshops} ($512 \times 512$) and Kodak24~\cite{KodakImages} ($128 \times 128$) datasets were used for each of the protocols. To assess the restoration qualities of ARKNet and XY-Deblur, input images were synthetically degraded using a $5 \times 5$ average blur kernel. 
As $Q$ is principally a structure feature which is derived from the Luma (Y) channel, only the Luma channel images were used during training and inference. 
To restore the blurry images, we used a hybrid loss as follows.

\begin{equation}
    \mathcal{L}_c = \mathcal{L}(I, \tilde{I}) - \beta \cdot Q(\tilde{I})  
    \label{eq:composite_loss}
\end{equation}

where $I$ and $\tilde{I}$ are the Ground Truth (GT) and restored images, respectively; $\mathcal{L}$ is the \textit{standard} loss used by the DNN models, and $\beta$ is the hyper-parameter which controls the amount of sharpness in the restored image. To produce a positive change in sharpness of the restored images, $\beta$ was set to $0.1$. During our ablation studies, we noted that a higher amount led to ringing artifacts and noise amplification. Visual examples of this phenomenon can be found in our supplementary materials.

\subsection{Protocols}
\label{sec:dataset_methods}
A total of four protocols were designed to study the effects of sharpness and sharpness-based losses on the Human Visual System (HVS). 

{\noindent \textbf{$P_1$: Method of Adjustment.}} To gauge the preferred level of sharpness for providing estimates of optimal sharpness across different image contents, subjects viewed a single image equipped with a slider and were tasked with adjusting the slider to their preferred level of sharpness. The slider was initialized at 0, which corresponded to a test image with no distortions or gains in sharpness. 

{\noindent \textbf{$P_2$: Two-Alternative Forced Choice (2AFC).}} To assess whether incorporating a composite loss produced sharper restorations, subjects were presented with pairs of images restored using different methods and loss functions. For each pair, they were forced to select the image they perceived as sharper.

{\noindent \textbf{$P_3$: Double Stimulus Impairment Scale (DSIS).}} To examine the correlation between standard Full-Reference (FR) and No-Reference (NR) metrics with DMOS, subjects were shown a Ground Truth (GT) image alongside a sharpened version, generated at varying intensity increments of 0.25 units using unsharp masking (\texttt{imsharpen()} in MATLAB). They were then asked to rate the quality of the sharpened image relative to the GT on a scale of 0 to 100, where a value of 50 meant both images were perceptually identical. A value below 50 corresponded to the sharpened image being \textit{worse} than GT, and a value above 50 meant that the sharpened image was \textit{better} than GT. This range was chosen over the standard 5-point scale to better capture perceptual differences. 
We acknowledge the deviation from standard protocol specifications as DSIS assumes that the test image will always be ranked at a lower score than the reference image, which is applicable for degradations such as blur, noise and JPEG artifacts. However, for sharpness enhancement, a slight sharpness increase over the reference image is usually perceptually better. This observation led us to modify the protocol to better suit our problem. 
    
{\noindent \textbf{$P_4$: Pairwise Comparison Evaluation}} To assess the overall quality of restorations rather than sharpness, subjects simultaneously viewed two restored images and rated both individually on a scale of 0 to 100, where higher scores indicated better quality.
\subsection{Participants and Data Pre-processing}
\label{sec:participants_processing}
The study involved 28 participants who provided their informed consent. SUREAL~\footnote{https://github.com/Netflix/sureal} was used to account for subject bias and inconsistency, in accordance with standards ITU-P.910. 
\subsection{Evaluation Metrics, IQA Databases and Correlation Coefficients}
For examining metrics which correctly take into consideration sharpness and artifacts produced by over-sharpening, we measured the DMOS on our dataset along with the KADID-10K~\cite{lin2019kadid}, which is the only dataset that contains such distortions, providing 1 reference image along with 5 different over-sharpened images, totaling 450 images of the class sharpness and contrast along with the corresponding DMOS score. However, unlike our experiment, the sharpness levels are neither uniformly incremented nor publicly available. 

As part of the suite of IQA metrics, we tested both No-Reference (NR) and Full-Reference (FR) metrics. For NR metrics, we selected $Q$~\cite{zhu2010automatic},
Natural Scene Statistics (NSS) based metrics, BRISQUE~\cite{mittal2012no} and a neural metric NIMA~\cite{talebi2018nima}. 
As part of Full-Reference (FR) metrics, we chose PSNR and its HVS variants, PSNR-HVS, PSNR-HVS-M~\cite{egiazarian2006new}, PSNR-HA and PSNR-HMA~\cite{ponomarenko2011modified}, as well as SSIM~\cite{1284395} and GSSIM~\cite{chen2006gradient}
which incorporates sharpness into SSIM. VIF~\cite{sheikh2006image} was also selected as it differentiates between a positive increase in sharpness and over-sharpness. Additionally, a neural metric LPIPS~\cite{zhang2018unreasonable} was also tested. 

To compute correlation between the above-mentioned metrics and DMOS, we compute the Pearson Linear Correlation Coefficient (PLCC), Kendall Rank Correlation Coefficient (KRCC) and Spearman's Rank Correlation Coefficient (SROCC). Following the VQEG guidelines~\cite{video2000final} of fitting quality metrics to perceptual scores, we employ a four-parameter logistic function to map metric values to DMOS. Our optimisation process combined exhaustive search for coarse tuning with Broyden-Fletcher-Goldfarb-Shanno (BFGS) for the refinement of the logistic fit parameters.

\section{Results and Discussion}
\label{sec:results}
\subsection{$Q$ as an indicator of preferred sharpness}
\label{subsec:Q_pref_sharpnes}
\begin{figure}
    \centering
    \includegraphics[width=1.1\linewidth]{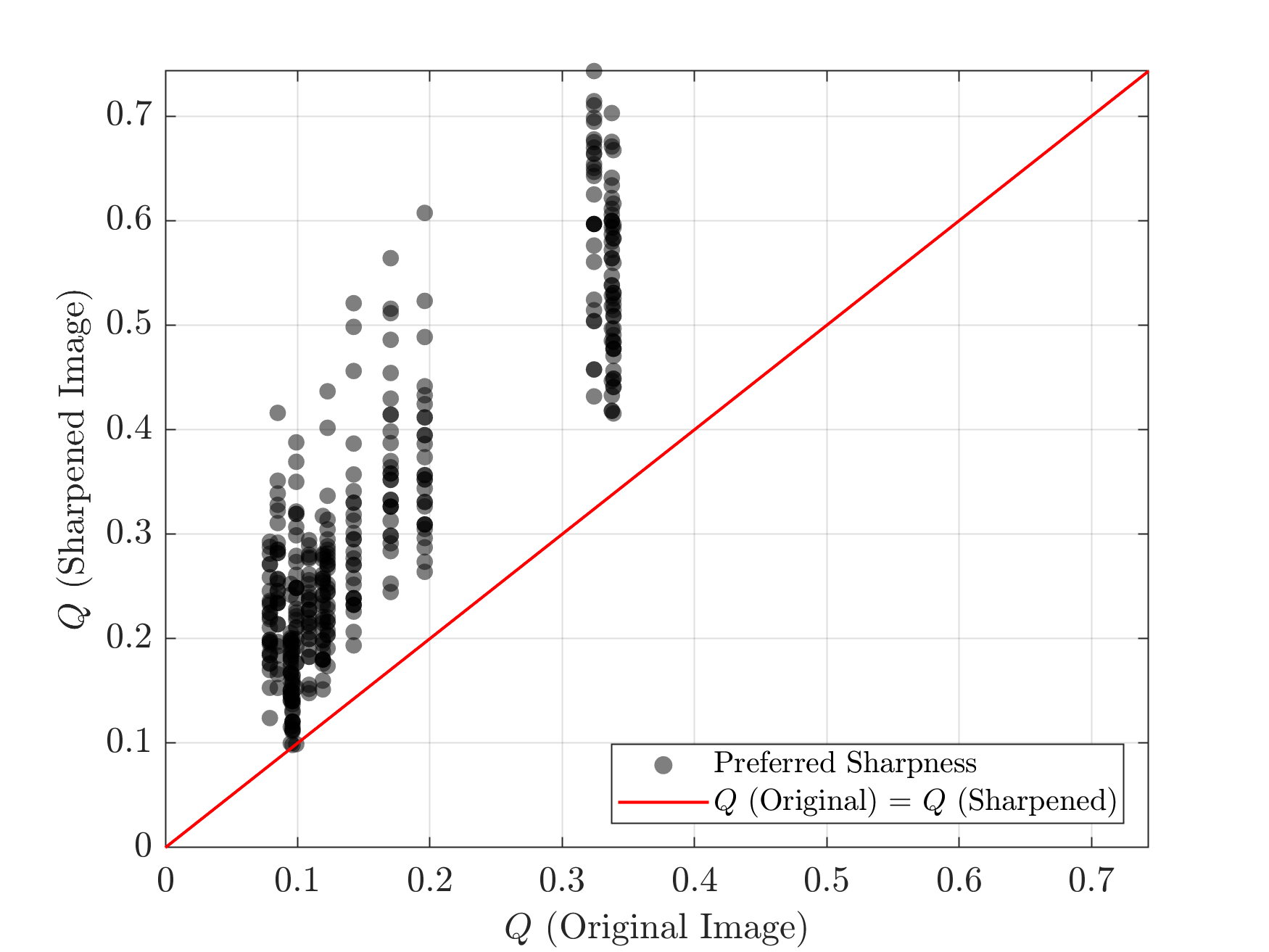}
    \caption{\textit{Sharpness ($Q$) Comparison of Reference Image and Sharpened Image}. Images were sharpened to their preferred sharpness levels, showing that subjects preferred sharper images overall, as compared to the reference image.}
    \label{fig:protocol1_scatter}
\end{figure}

It is a well-established fact that slightly over-sharpened images are preferred during a viewing experience. However, relatively few works examine this phenomenon with the help of IQA metrics. Through $P_1$, we used NR metric $Q$ to confirm this phenomenon. Fig.~\ref{fig:protocol1_scatter} shows the plot where the reference (GT) image was sharpened to preferred sharpness amounts, producing an enhanced image. A total of 15 images were sharpened based on 28 user preferences for each image. The number of images was chosen primarily for practical reasons to minimize participant fatigue during the study. We then plot sharpness ($Q$) of GT image on the x--axis and sharpness of  enhanced image on the y--axis. The red line denotes the scenario when both the enhanced and GT image have the same sharpness. We see that the preferred sharpness of enhanced images is above the line, which confirms that for varying images (in terms of textural details), subjects preferred sharper images, overall. 

Furthermore, we noted that preferred sharpness amounts for GT images which already had high frequency components (edges, corners) was lower than those with mostly flat patches, thus demonstrating that $Q$ can also be used as an indicator of \textit{preferred sharpness}, in addition to overall sharpness. This experiment demonstrated that deblurring models must focus on explicitly producing slightly sharper restorations, in addition to restoring lost frequencies from the blurry image. 

\subsection{Using $Q$ to produce visually sharper and aesthetically better images}
\label{subsec:Q_visually_sharper}

\begin{figure}
    \centering
    \begin{tabular}{cc}
    \includegraphics[width=0.45\linewidth]{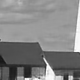} &  
      \includegraphics[width=0.45\linewidth]{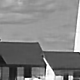} \\
    (a) Restored with $\mathcal{L}$ & (b) Restored with $\mathcal{L}_c$
    \end{tabular}
    \caption{\textit{Example of an image restored with composite loss $\mathcal{L}_c$} (SI-PSNR = 27.26) \textit{and standard loss $\mathcal{L}$} (SI-PSNR = 35.15). Such images were shown to subjects within the framework of Protocol 2, wherein they were asked to select the sharper of the two.}
    \label{fig:protocol2_image_comp}
    \vspace{-1em}
\end{figure}

To answer the question whether using a composite loss (Eqn.~\ref{eq:composite_loss}) which included $Q$, produced sharper images after deblurring, subjects were shown two images (Fig.~\ref{fig:protocol2_image_comp}), restored using a standard loss $\mathcal{L}$ that did not include $Q$ (Fig.~\ref{fig:protocol2_image_comp} (a)) and the other restored with the composite loss $\mathcal{L}_c$ (Fig.~\ref{fig:protocol2_image_comp} (b)). Around $78\%$ of the participants selected the image restored with $\mathcal{L}_c$ as being sharper than those restored with $\mathcal{L}$, thus showing that incorporating $Q$ as part of the loss indeed produces sharper restorations. 

\begin{figure}
    \centering
    \includegraphics[width=0.7\linewidth]{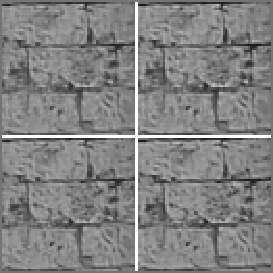}
    \caption{\textit{Example of restorations from different DNN-based deblurring mechanisms with different losses}. (Top, Bottom): Images restored using ARKNet~\cite{aurangabadkar2024sharpness} and XY-Deblur~\cite{ji2022xydeblur}. (Left, Right): Images restored using $\mathcal{L}$ and $\mathcal{L}_c$. Images restored with $\mathcal{L}_c$ from both deblurring methods produce slightly sharper edges, resulting in aesthetically better images.}
    \label{fig:protocol4_montage}
\end{figure}

Let us now discuss whether this approach leads to aesthetically \textit{better} images. The Mean Opinion Score (MOS) of images restored using  $\mathcal{L}_c$ was higher than the images restored using $\mathcal{L}$ ($56.78$ and $55.24$, respectively) (see Fig.~\ref{fig:protocol4_montage}). Performing a paired t-test on the metric values, we found that the observed difference between means of metrics of images restored using $\mathcal{L}_c$ as opposed to $\mathcal{L}$ were not significant at the 5\% level ($p < 0.05$). To better understand pairwise preferences, we assign 1 to an image with a higher MOS and a 0 otherwise. From this binarised analysis, we see that images restored with $\mathcal{L}_c$ were preferred in $67\%$ of the comparisons. 

\begin{figure}
    \centering
    \includegraphics[width=1.0\linewidth]{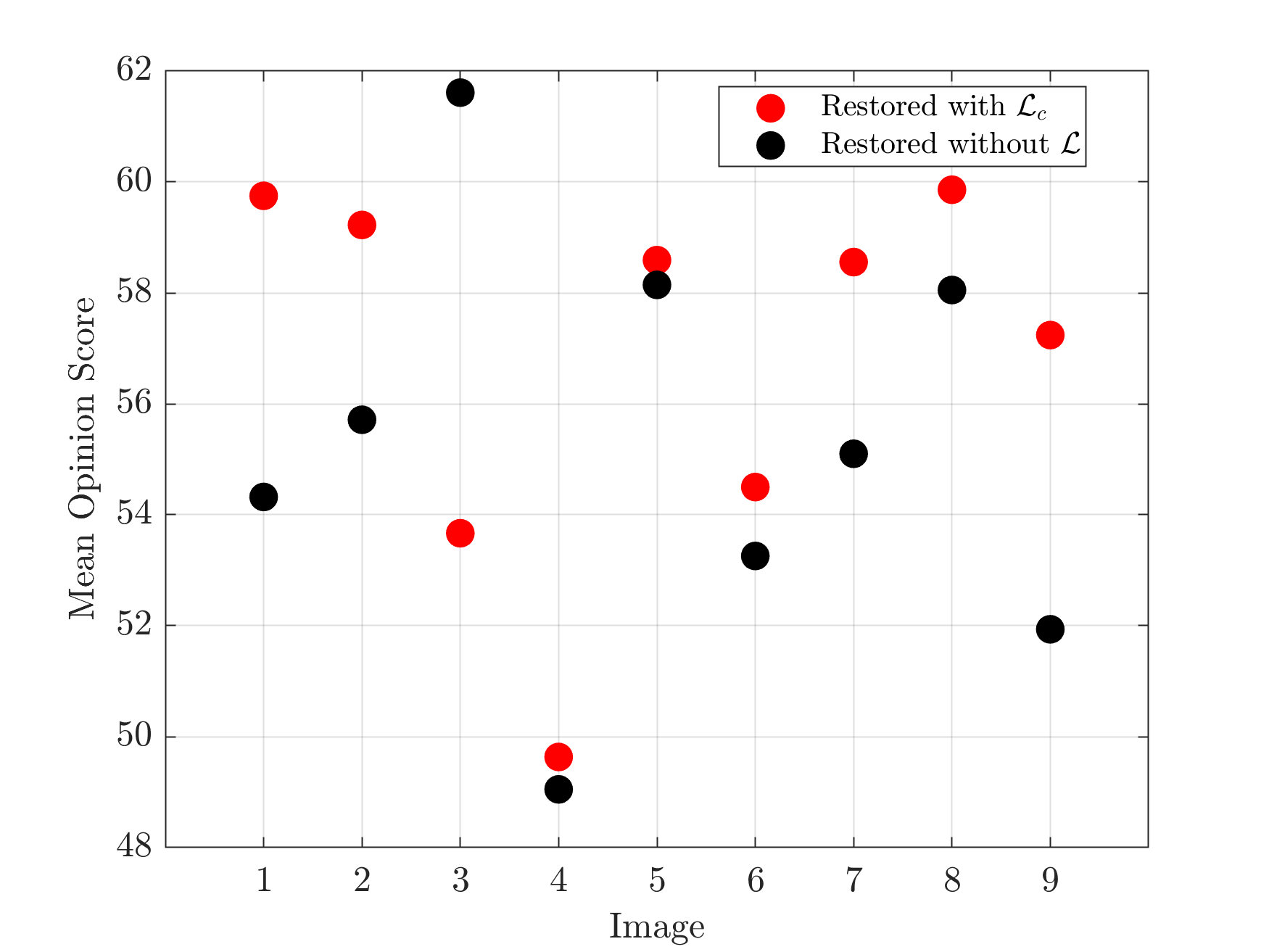}
    \caption{\textit{Comparison of MOS for images restored with separate losses averaged over two separate DNN models: ARKNet and XY-Deblur}. Red dots higher than black dots denote that subjects preferred images restored with $\mathcal{L}_c$ rather than $\mathcal{L}$, showing that subjects preferred slightly sharper restorations.}
    \label{fig:protocol4_observerWise_scores}
\end{figure}

Figure~\ref{fig:protocol4_observerWise_scores} shows the observer-wise MOS per image averaged over two methods. The MOS of images restored using $\mathcal{L}_c$ (highlighted in red) is slightly higher than the MOS of images restored using $\mathcal{L}$, thus showing that deploying $Q$ as a composite loss produces \textit{perceptually better} images.

\subsection{Correlation between IQA Metrics and Sharpness}
\label{subsec:correlation_iqa_metrics}

\begin{table*}[]
\centering
\resizebox{\textwidth}{!}{%
\begin{tabular}{@{}cccccc
>{\columncolor[HTML]{D9D9D9}}c 
>{\columncolor[HTML]{D9D9D9}}c 
>{\columncolor[HTML]{D9D9D9}}c 
ccccccc@{}}
\toprule
\textit{\textbf{Dataset}}              & \textit{\textbf{}}      & \textbf{SSIM}~\cite{1284395}                  & \textbf{LPIPS}~\cite{zhang2018unreasonable}                                        & \textbf{VIF}~\cite{sheikh2006image}                  & \textbf{GSSIM}~\cite{chen2006gradient}                & \textbf{$Q$}~\cite{zhu2010automatic}                    & \textbf{BRISQUE}~\cite{mittal2012no}              & \textbf{NIMA}~\cite{talebi2018nima}                & \textbf{PSNR}                 & \textbf{PSNR-HVS-M}~\cite{egiazarian2006new}           & \textbf{PSNR-HMA}~\cite{ponomarenko2011modified}             & \textbf{SI-PSNR}               & \textbf{SI-SSIM}               \\ \midrule
                                       & \textit{\textbf{PLCC}}  & 0.885  & {\color[HTML]{FE6100} \textbf{0.921}} & 0.791 & {\color[HTML]{648FFF} \textbf{0.909}} & 0.462 & 0.568 & 0.135 & 0.876 & 0.873 & 0.875 & \underline{{\textit{\textbf{0.905}}}} & \underline{\textit{\textbf{0.886}}}                \\
                                       & \textit{\textbf{KRCC}}  & 0.702  & {{0.741}} & 0.600 & 0.706 &  0.320 & 0.405 & 0.098 & 0.696 & 0.685 & 0.680 & \underline{\textbf{\textit{0.724}}} & \underline{\textit{\textbf{0.672}}}                \\
\multirow{-3}{*}{\textbf{KADID-10K}~\cite{lin2019kadid}}   & \textit{\textbf{SROCC}} & 0.887  & {{0.915}} & 0.794 & 0.893 & 0.461 & 0.582 & 0.119 & 0.877 & 0.871 & 0.866 & \underline{\textit{\textbf{0.901}}} & \underline{\textit{\textbf{0.871}}}                \\ \midrule
                                       & \textit{\textbf{PLCC}}  & \color[HTML]{FE6100} \textbf{0.912} & 0.855                                                & 0.720                        & 0.887                        & 0.536                        & 0.624                        & 0.315                        & 0.720                        & 0.761                        & 0.702                        & \underline{\textit{\textbf{0.857}}}                & { \color[HTML]{648FFF}\underline{\textit{\textbf{0.910}}}} \\
                                       & \textit{\textbf{KRCC}}  & {0.753} & 0.686                                                & 0.545                        & 0.712                        & 0.381                        & 0.432                        & 0.146                        & 0.554                        & 0.589                        & 0.540                        & \underline{\textit{\textbf{0.681}}}                & \underline{\textit{\textbf{0.758}}} \\
\multirow{-3}{*}{\textbf{Our Dataset}} & \textit{\textbf{SROCC}} & {0.920} & 0.870                                                & 0.728                        & 0.900                        & 0.519                        & 0.612                        & 0.163                        & 0.740                        & 0.783                        & 0.726                        & \underline{\textit{\textbf{0.869}}}                & \underline{ \textit{\textbf{0.923}}} \\ \bottomrule
\end{tabular}%
}
\caption{\textit{Correlation between IQA metrics and DMOS over two separate datasets}. The columns highlighted in grey are NR-IQA. The best and second best performing metrics are highlighted in orange and blue respectively. Our proposed metrics are in italics, bold and underlined (cf. last two columns). Our proposed SI-PSNR produces the best correlation in comparison to all other PSNR variants.}
\label{tab:protocol3_table_variant2}
\end{table*}

Table~\ref{tab:protocol3_table_variant2} shows that LPIPS outperforms all other metrics in terms of PLCC on the KADID-10K dataset. The proposed Sharpness Informed PSNR (SI-PSNR) (underlined and in bold) achieves a PLCC of 0.905. LPIPS is a neural metric which extracts features from DNN layers to predict image quality, making it complex to compute, whereas SI-PSNR is independent of such nuances, thus making it easier to compute on resource-constrained machines. On our dataset, once more, SI-PSNR is better than any other variant of PSNR. On the same dataset, however, SSIM appears to be the best in terms of correlations with DMOS, although in comparison with our metric, Sharpness Informed SSIM (SI-SSIM) performs nearly as well, with a PLCC of about 0.0013 units less than SSIM. Evidence for this is available in our supplementary material. It is interesting to note that VIF, a metric used as a component in VMAF~\footnote{https://netflixtechblog.com/toward-a-practical-perceptual-video-quality-metric-653f208b9652}, gave a PLCC of approximately 0.75 on average, which had the lowest correlation for FR metrics. Our results suggest that SI-PSNR may be deployed in production pipelines as a low-cost metric which correlates well with sharpness and overall quality. We note that SSIM correlates well and may be used as an alternative. 

\section{Conclusion}
\label{sec:conclusion}
We conducted a subjective study to understand the effects of sharpness on the Human Visual System and the ability of sharpness-based loss functions deployed in DNN-based deblurring mechanisms. We showed that overall, subjects preferred sharp images, as compared to the reference image. We introduced a novel dataset of images with controlled sharpness along with DMOS. We proposed a novel class of \textit{Sharpness Informed} (SI) metrics which combine classical metrics (PSNR, SSIM) along with $Q$ to correctly penalize ringing. These metrics correlate well with sharpened images (e.g, more than 0.88 in PLCC for SI-SSIM) and outperform variants of PSNR, VIF and neural metrics such as LPIPS in terms of PLCC on two separate datasets. 

In addition to this, we also showed that images restored using losses which explicitly target sharpness are preferred in $67\%$ of binarized comparisons. Furthermore, $78\%$ of the subjects said that using a composite loss with $Q$ indeed produced visibly sharper images. Our future work consists of deploying the proposed SI-PSNR as a loss function in deblurring models.

\bibliographystyle{IEEEbib}
\bibliography{refs}

\begin{thebibliography}{10}

\bibitem{kupyn2018deblurgan}
O.~Kupyn, V.~Budzan, M.~Mykhailych, D.~Mishkin, and J.~Matas,
\newblock ``Deblurgan: Blind motion deblurring using conditional adversarial networks,''
\newblock in {\em Proceedings of the IEEE conference on CVPR}, 2018, pp. 8183--8192.

\bibitem{he2024dual}
Yuhang He, Senmao Tian, Jian Zhang, and Shunli Zhang,
\newblock ``Dual attention enhanced transformer for image defocus deblurring,''
\newblock in {\em 2024 IEEE International Conference on Image Processing (ICIP)}. IEEE, 2024, pp. 1487--1493.

\bibitem{1284395}
Z.~Wang, A.C. Bovik, H.R. Sheikh, and E.P. Simoncelli,
\newblock ``Image quality assessment: from error visibility to structural similarity,''
\newblock {\em IEEE Transactions on Image Processing}, vol. 13, no. 4, pp. 600--612, 2004.

\bibitem{egiazarian2006new}
K.~Egiazarian, K.~Astola, N.~Ponomarenko, V.~Lukin, F.~Battisti, and M.~Carli,
\newblock ``New full-reference quality metrics based on hvs,''
\newblock in {\em Proceedings of the second international workshop on video processing and quality metrics}, 2006.

\bibitem{zhang2018unreasonable}
R.~Zhang, P.~Isola, A.~Efros, E.~Shechtman, and O.~Wang,
\newblock ``The unreasonable effectiveness of deep features as a perceptual metric,''
\newblock in {\em Proceedings of the IEEE conference on CVPR}, 2018, pp. 586--595.

\bibitem{aurangabadkar2024sharpness}
U.~Aurangabadkar, D.~Ramsook, and A.~Kokaram,
\newblock ``A sharpness based loss function for removing out-of-focus blur,''
\newblock in {\em 2024 IEEE 26th International Workshop on Multimedia Signal Processing (MMSP)}. IEEE, 2024, pp. 1--6.

\bibitem{zhu2010automatic}
X.~Zhu and P.~Milanfar,
\newblock ``Automatic parameter selection for denoising algorithms using a no-reference measure of image content,''
\newblock {\em IEEE transactions on image processing}, vol. 19, no. 12, pp. 3116--3132, 2010.

\bibitem{zamir2021multi}
Syed~Waqas Zamir, Aditya Arora, Salman Khan, Munawar Hayat, Fahad~Shahbaz Khan, Ming-Hsuan Yang, and Ling Shao,
\newblock ``Multi-stage progressive image restoration,''
\newblock in {\em Proceedings of the IEEE/CVF conference on computer vision and pattern recognition}, 2021, pp. 14821--14831.

\bibitem{krasula2017quality}
L.~Krasula, P.~Le~Callet, K.~Fliegel, and M.~Kl{\'\i}ma,
\newblock ``Quality assessment of sharpened images: Challenges, methodology, and objective metrics,''
\newblock {\em Transactions on Image Processing}, vol. 26, pp. 1496--1508, 2017.

\bibitem{mittal2012no}
A.~Mittal, A.~K. Moorthy, and A.~C. Bovik,
\newblock ``No-reference image quality assessment in the spatial domain,''
\newblock {\em IEEE Transactions on image processing}, vol. 21, no. 12, pp. 4695--4708, 2012.

\bibitem{11226201}
Uditangshu Aurangabadkar, Darren Ramsook, and Anil Kokaram,
\newblock ``Impact of a sharpness based loss function for removing out-of-focus blur,''
\newblock in {\em 2025 33rd European Signal Processing Conference (EUSIPCO)}, 2025, pp. 601--605.

\bibitem{ponomarenko2015image}
N.~Ponomarenko et~al.,
\newblock ``Image database tid2013: Peculiarities, results and perspectives,''
\newblock {\em Signal processing: Image communication}, vol. 30, pp. 57--77, 2015.

\bibitem{sheikh2006statistical}
H.R. Sheikh, M.F. Sabir, and A.C. Bovik,
\newblock ``A statistical evaluation of recent full reference image quality assessment algorithms,''
\newblock {\em IEEE Transactions on image processing}, vol. 15, no. 11, pp. 3440--3451, 2006.

\bibitem{bt2019methodologies}
ITU BT,
\newblock ``Methodologies for the subjective assessment of the quality of television images,''
\newblock {\em Document Recommendation ITU-R BT}, pp. 500--14, 2023.

\bibitem{ji2022xydeblur}
S.~Ji, J.~Lee, S.~Kim, J.~Hong, S.~Baek, S.~Jung, and S.~Ko,
\newblock ``Xydeblur: Divide and conquer for single image deblurring,''
\newblock in {\em Proceedings of the IEEE/CVF conference on CVPR}, 2022, pp. 17421--17430.

\bibitem{Agustsson_2017_CVPR_Workshops}
E.~Agustsson and R.~Timofte,
\newblock ``Ntire 2017 challenge on single image super-resolution: Dataset and study,''
\newblock in {\em The IEEE Conference on CVPR Workshops}, July 2017.

\bibitem{KodakImages}
``Kodak lossless true color image suite. [online]. available: http://r0k.us/graphics/kodak/,'' 2007.

\bibitem{lin2019kadid}
H.~Lin, V.~Hosu, and D.~Saupe,
\newblock ``Kadid-10k: A large-scale artificially distorted iqa database,''
\newblock in {\em 2019 Eleventh International Conference on Quality of Multimedia Experience (QoMEX)}. IEEE, 2019, pp. 1--3.

\bibitem{talebi2018nima}
H.~Talebi and P.~Milanfar,
\newblock ``Nima: Neural image assessment,''
\newblock {\em IEEE transactions on image processing}, vol. 27, no. 8, pp. 3998--4011, 2018.

\bibitem{ponomarenko2011modified}
N.~Ponomarenko, O.~Ieremeiev, V.~Lukin, K.~Egiazarian, and M.~Carli,
\newblock ``Modified image visual quality metrics for contrast change and mean shift accounting,''
\newblock in {\em 11th International Conference The Experience of Designing and Application of CAD Systems in Microelectronics (CADSM)}. IEEE, 2011, pp. 305--311.

\bibitem{chen2006gradient}
G.~Chen, C.~Yang, and S.~Xie,
\newblock ``Gradient-based structural similarity for image quality assessment,''
\newblock in {\em International Conference on Image Processing}. IEEE, 2006, pp. 2929--2932.

\bibitem{sheikh2006image}
Hamid~R Sheikh and Alan~C Bovik,
\newblock ``Image information and visual quality,''
\newblock {\em IEEE Transactions on image processing}, vol. 15, no. 2, pp. 430--444, 2006.

\bibitem{video2000final}
Video Quality~Experts Group et~al.,
\newblock ``Final report from the video quality experts group on the validation of objective quality metrics for video quality assessment,''
\newblock {\em see http://www. its. bldrdoc. gov/vqeg/projects/frtv phaseI}, 2000.

\end{thebibliography}

\end{document}